# Review on ferroelectric/polar metals

W. X. Zhou[1] and A. Ariando[1,2,3] *

[1]NUSNNI-NanoCore, National University of Singapore, Singapore 117411, Singapore

[2]Department of Physics, National University of Singapore, Singapore 117542, Singapore

[3]NUS Graduate School for Integrative Sciences and Engineering, National University of Singapore, Singapore 117456, Singapore.

To whom correspondence should be addressed: ariando@nus.edu.sg

**The possibility of reconciliation between seemingly mutually exclusive properties in one system can not only lead to theoretical breakthroughs but also potential novel applications. The research on the coexistence of two purportedly contra-indicated properties, ferroelectricity/polarity and conductivity, proposed by Anderson and Blount over 50 years ago was recently revitalized by the discovery of the first unambiguous polar metal $LiOsO_3$ and further fueled by the demonstration of the first switchable ferroelectric metal $WTe_2$. In this review, we first discuss the reasons why the coexistence of ferroelectricity/polarity and conductivity have been deemed incompatible, followed by a review on the history of ferroelectric/polar metals. Secondly, we review the important milestones along with the corresponding mechanisms for the ferroelectric/polar metallic phases in these materials. Thirdly, we summarize the design approaches for ferroelectric/polar metals. Finally, we discuss the future prospects and potential applications of ferroelectric/polar metals.**

## 1. Introduction

Incorporating multiple physical properties, such as ferromagnetism, ferroelectricity and (super-) conductivity into one system is one promising alternative, among others, to sustain the continuous advancements in electronics beyond Moore[1-4]. However, some pairs of properties were/are considered to be mutually exclusive in fundamental physics. Over the past decade or so, the exploration of the coexistence of seemingly mutually exclusive properties has generated a flurry of interest, as its discovery is not only crucial for potential applications but also for fundamental physics. For example, ferromagnetism and ferroelectricity were deemed mutually exclusive because ferromagnetism normally requires partially filled $d$ orbitals, while ferroelectricity prefers empty $d$ orbitals[5]. However, this perception is overturned by the recent discovery of their coexistence in $(Ca_ySr_{1-y})_{1.15}Tb_{1.85}Fe_2O_7$[6] and at the Fe/BaTiO$_3$ interface[7]. Another ground-breaking example is the discovery of the coexistence of ferromagnetism and superconductivity in iron-based superconductors and at the interface of LaAlO$_3$/SrTiO$_3$, although ferromagnetism is expected to destroy the pairing interaction responsible for superconductivity[8, 9]. Recently, another pair of contra-indicated properties – ferroelectricity/polarity and conductivity – has strongly stirred the research community.

Based on inversion symmetry, a crystal structure can be divided into centrosymmetric and non-centrosymmetric[10]. A centrosymmetric structure is non-polar. A non-centrosymmetric structure can be further divided into polar and non-polar. The ordered polar electric dipoles give rise to ferroelectricity, whose polarization can be switched with an external electric field. A ferroelectric undergoes a phase transition from non-polar paraelectric state into a polar ferroelectric state upon cooling. In conventional displacive ferroelectrics, such as BaTiO$_3$ (BTO), this phase transition is characterized by a decreasing frequency of a transverse optical phonon mode (the soft mode) which drops to zero at the transition point and then becomes imaginary in the ferroelectric phase, corresponding to a collective displacement of ions from their centrosymmetric positions[11, 12]. The ferroelectric instability is explained by the delicate balance between long-range Coulomb forces favoring the polar ferroelectric phase and the short-range repulsions favoring the non-polar paraelectric phase[13]. In addition, it has been shown that the covalent hybridization between Ti $3d$ and O $2p$ orbitals is required to weaken the short-range repulsions to stabilize the ferroelectric phase[5, 12, 13]. Introducing itinerant charge carriers into a ferroelectric insulator is expected to screen the long-range Coulomb forces and quench ferroelectricity. That is the reason why ferroelectricity and conductivity are not expected to coexist in one material. Despite of this constraint, scientists have worked diligently to develop a ferroelectric metal to break through the limit.

## 2. A brief history of polar and ferroelectric metals

Here, we present a brief history of the trial and error process in search for ferroelectric metals. In 1965, Anderson and Blount first proposed the possibility of a ferroelectric metal providing that the itinerant electrons do not interact strongly with the transverse optical phonons[14]. To be considered a ferroelectric metal, four criteria have to be met: 1) a second-order phase transition, 2) the removal of an inversion symmetry, 3) the appearance of a polar axis and 4) the demonstration of polarization switchability[15,16]. For a polar or ferroelectric-like metal, the former three criteria have to be met. We note that until around 2016, the subtle differences of the terms "polar metal", "ferroelectric-like metal" and "ferroelectric metal" are still quite obscure and the research community often used them interchangeably[17]. This is probably because of two reasons: (i) In a typical paraelectric-to-ferroelectric phase transition, the term "ferroelectric" is almost always associated with the term "polar" and the term "paraelectric" is closely tied to "non-polar"; (ii) In the modern theory of ferroelectricity and polarization, only the change in polarization under electric field switching is physically meaningful and since the electric field cannot exist in a metal due to electrostatic screening[17, 18], it was widely believed that a polarization and hence ferroelectricity could not be defined in a metal. However, with the discovery of the truly electric-field-switchable ferroelectric metal WTe$_2$[19, 20], it is now not appropriate to use these terms interchangeably. In this article, we will try our best to use the terms as scientifically rigid as possible, specifically we will use the terms "ferroelectric" and "polarization" to describe "ferroelectric metals" with switchable polarization and the terms "polar", "ferroelectric-like" and "polarity" to describe the materials whose switchability has yet to be demonstrated. Nevertheless, as a review article, we will inevitably encounter cases where we need to cite results from earlier reports in which the terms were not well differentiated. In those cases, we would put double quotes around the terms where we believe they were misused. Initial research was focused on V$_3$Si (and other A15-type superconductors), which adopts the cubic A15 or β-W structure above 21 K, below which it undergoes a transition to a tetragonal phase[21]. Earlier experiments showed the transition was second order and it was not possible to describe the transition in V$_3$Si using strain as the only order parameter, and hence Anderson and Blount predicted V$_3$Si to be a "ferroelectric" metal[14]. However, detailed studies showed the structural transitions to be weakly first order and that strain was indeed the appropriate order parameter[22-24]. Since then, there was no experimental evidence of "ferroelectric" metal for almost four decades until 2004 when Sergienko et al. reported Cd$_2$Re$_2$O$_7$ as a promising candidate[25]. However, it was later found that although the lowest-symmetry phase is non-centrosymmetric, it is nonpolar and is better described as a piezoelectric metal rather than a "ferroelectric" metal[26]. In 2010, Kolodiazhnyi reported another promising "ferroelectric" metal, electron-doped BaTiO$_{3-\delta}$[27,28], where the low-symmetry "ferroelectric" phases were found to persist up to a critical electron concentration of $1 \times 10^{20}$ cm$^{-3}$. However, later structural measurements suggested that the "ferroelectric" ordering and conductivity are separated into two phases and thus do not coexist microscopically[29]. After nearly half a century's search, the first unambiguous polar metal LiOsO$_3$ was discovered by Shi et al. in 2013[15]. Based on detailed structural and transport experiments, Shi et al. confirmed that metallic LiOsO$_3$ undergoes a continuous second-order centrosymmetric ($R\bar{3}c$) to non-centrosymmetric and polar ($R3c$) phase transition at 140 K that is structurally equivalent to the ferroelectric transition of LiNbO$_3$. In 2014, Puggioni and Rondinelli proposed an operational model for designing polar metals, coined "decoupled electron mechanism" (DEM) model, where it expands the Anderson-Blount model and states that the existence of polar metals "relies on weak coupling between the electrons at the Fermi level, and the (soft) phonon(s) responsible for removing inversion symmetry"[30]. This DEM model is later found to be useful in explaining the stability of many polar metals. Although the discovery of a solid Anderson-Blount type polar metal greatly enhances our understanding of the nature of ferroelectricity and conductivity, the core functionality of ferroelectricity is still missing – the switchablity. In 2016, Filippetti and coworkers' theoretical paper predicted an electric-field-switchable ferroelectric metal Bi$_5$Ti$_5$O$_{17}$, which can sustain a sizable potential drop along the polar direction, as needed to reverse its polarization by an external bias despite being a metal[31]. However, the first experimental realization of a truly ferroelectric metal WTe$_2$ is only discovered very recently in 2018[19, 20].

## 3. Materials

Over the past decades, many types of polar and ferroelectric metals have been discovered. In ref. 17, Benedek and Birol listed around 70 polar metals from a non-exhaustive search of the Inorganic Crystal Structure Database (ICSD)[17]. Here, we list more types of polar metals that have been reported (or predicted) based on our knowledge in Table 1. As can be seen from the table, the research was mainly focused on oxides and various two-dimensional materials. Hence, in the following section, we will focus primarily on oxide-based and two-dimensional ferroelectric/polar metals.

Table 1. A list of reported/predicted ferroelectric/polar metals

| Types | Materials and references |
|---|---|
| Perovskite oxides | $BaTiO_{3-\delta}$[12, 27, 28]; La-doped $BaTiO_3$[32], $BaTiO_3/SrTiO_3/LaTiO_3$[33], $LaAlO_3/Ba_{0.2}Sr_{0.8}TiO_3/SrTiO_3$[34], $PbTiO_{3-\delta}$[35], $PbTi_{1-x}Nb_xO_3$[36], $CaTiO_{3-\delta}$[17]; $BiFeO_{3-\delta}$[37], $Sr_{1-x}Ca_xTiO_{3-\delta}$[38], n-doped $BaMnO_3$[39], n-dope $BiAlO_3$[39], n-doped $LaFeO_3$/ $YFeO_3$[39], $NdNiO_3$[40], $LiOsO_3$[15, 41, 42], $LiNbO_3$[43], $MgReO_3$[41], $TiGaO_3$[41] |
| Layered perovskites | $Ca_3Ru_2O_7$[44]; $(Sr,Ca)Ru_2O_6$[30], $Bi_5Ti_5O_{17}$[31] ; $Cd_2Re_2O_7$[25], $BiPbTi_2O_6$[45], $La_2Ti_2O_7$[39], $Sr_2Nb_2O_7$[39], $Ca_3Ti_2O_7$[39] |
| Antiperovskites | Strained $ACNi_3$ (A = Mg, Zn, and Cd)[46], $CeSiPt_3$[47] |
| LiGaGe-type structure | $LiGaGe$[48], $SrHgPb$[49], $SrHgSn$[49], $CaHgSn$[49], $KMgSb_{0.2}Bi_{0.8}$[50], $CaAgBi$[51], $LiZnBi$[52], $LaAuGe$[53], $LaPtSb$[53] |
| Two-dimensional materials | $WTe_2$[19, 20], $MoTe_2$[54], $CrN$[55], $CrB_2$[55] |
| Others | group-V elements (P, As, Sb, Bi)[56], $SnP$[57], $BeAu$[58] |

## 4. Milestones

In this section, we discuss some materials that have played significant roles in the pursuit of ferroelectric/polar metals. We also review the theoretical progress in explaining the mechanisms of the coexistence of metallicity and ferroelectricity/polarity in these materials.

### 4.1 BaTiO$_3$ and other perovskite oxides

As a prototypical ferroelectric with a simple $ABO_3$ formula, BTO has played a significant role in understanding the origins of ferroelectricity as well as polar metals. In its bulk form, BTO is an insulator with a bandgap of 3.2 eV[59]. It undergoes a series of phase transitions from high-temperature paraelectric cubic to ferroelectric tetragonal, orthorhombic and rhombohedral phases at 403, 287 and 197 K, respectively[27]. The paraelectric-to-ferroelectric phase transition at the 403 K Curie temperature ($T_c$) is characterized by the evolution of the transverse optical (TO) $F_{1u}$ soft phonon mode[60]. Although the electric transport properties has been studied for decades in semiconducting *n*-type BTO[61], it was not until 2008 that Kolodiazhnyi et al. reported the metallic behavior in *n*-type BTO when the carrier density exceeds $10^{20}$ cm$^{-3}$ by oxygen vacancy doping[28]. Based on the observation that the low-symmetry phases of *n*-type BTO are preserved up to a critical carrier concentration of $n_c \approx 1.9 \times 10^{21}$ cm$^{-3}$, Kolodiazhnyi et al. proposed *n*-type BTO as a "ferroelectric" metal (Fig. 1a)[27]. This claim is supported by the theoretical work by Wang et al., who found that the polar displacements and the soft phonon mode in *n*-type BTO can be sustained up to a critical carrier density of 0.11 electron per unit cell (e/uc)[12], which is around $2 \times 10^{21}$ cm$^{-3}$. The persistence of polar distortions is attributed to a short-range portion of the Coulomb force (~ 5Å) with an interaction range of the order of the lattice constant, i.e. the Coulomb force is incompletely screened in this range[12]. However, Zhao et al. recently pointed out that this short-range interaction is not a portion of the Coulomb force but a force arising from the so-called "meta-screening effect", which is triggered by local lattice response accommodating the screening electrons[39]. Furthermore, this "meta-screening effect" is proposed to be largely universal in other ferroelectric oxides, such as PbTiO$_3$ (PTO) and BiFeO$_3$ (BFO)[39]. As noted above, the polarity and conductivity in bulk *n*-type BTO has been found to exist in different phases by neutron scattering measurements[29]. However, recent scanning transmission electron microscopy (STEM) measurements reveal that polarity and conductivity can coexist in a single phase in "ferroelectric" Ba$_{0.2}$Sr$_{0.8}$TiO$_3$ thin films[34] (Fig. 1b-d) and proximity-induced "ferroelectric" SrTiO$_3$ thin films[33]. In addition, the critical electron density in BTO can be increased by compressive strain, improving the functionality of BTO-based polar metals[62].

While the ferroelectricity in BTO is stabilized by the long-range Coulomb forces and the hybridization between Ti 3$d$ and O 2$p$ orbitals, there is another driving force for the ferroelectricity in another class of ferroelectrics, which involves the displacement of the A-site atoms in addition to/instead of that of the B-site atoms as in BTO[13, 63, 64]. For example, in PTO, in addition to the Ti displacements, the Pb atoms are also displaced and contribute a significant portion to the total polarization[13, 63, 64]. The Pb displacement is due to hybridization between the Pb (6$s$, 6$p$) bands and O 2$p$ bands, which reduces the short-range repulsions and enhances polarization[13, 63, 64]. This hybridization is called the lone-pair mechanism. It has been found that although itinerant electrons can effectively screen the long-range Coulomb forces, the lone-pair-driven A-site polar distortion is not strongly affected or even enhanced by electron doping[35]. This is because the electronic states corresponding to the lone-pair mechanism are away from the Fermi energy if electrons are doped, and the bottom of conduction bands in these materials are often the B-site states[35]. Thus the doping of electrons can be seen as a selective enlargement of the B-site ion radius, which stretches the A-O bonds[35]. This lone-pair-driven persistence of polarity has also been found in BFO[37]. It may also be applicable to other lone-pair-driven ferroelectrics, such as PbVO$_3$[65], SnTiO$_3$[66] and BiMnO$_3$[67].

Although the polar distortions in both A-site and B-site perovskites discussed above can be sustained upon itinerant carriers doping, they originate from electrostatic forces, which are susceptible to electrostatic screening by itinerant charge carriers. So one may ask if there is a non-electrostatic mechanism to sustain the polar distortions. Next we review progress in another type of perovskite polar metal due to a (non-electrostatic) geometric-lattice mechanism, which is responsible for the polar metallic state found in layered perovskite (Sr,Ca)Ru$_2$O$_6$[30] and Ca$_3$Ru$_2$O$_7$[44]. This geometric-lattice mechanism is similar to hybrid improper or trilinear coupling ferroelectricity.

Here, we briefly introduce the difference between proper and improper ferroelectricity, interested readers can find comprehensive discussions in refs. 68-71. Based on Landau theory of phase transitions, a proper ferroelectric, such as BTO and PTO, has its electrical polarization as the primary order parameter[70, 71]. Furthermore, the paraelectric-ferroelectric phase transition is typically described by zone-center soft phonons[69]. In contrast, an improper ferroelectric is one which the primary order parameter is not electrical polarization, but a quantity having another physical meaning and possessing other transformation properties, i.e. the polarization does not drive the paraelectric-ferroelectric transition but instead is a "slave" to some other primary order parameter[70, 71]. In addition, the paraelectric-ferroelectric phase transition is typically described by coupled zone-boundary phonons instead of zone-center phonons[69].

In materials that display a proper ferroelectric transition such as BTO, the free energy can be expanded in terms of the polarization[71]:

$$\mathcal{F}(P) = 1/2\alpha P^2 + 1/4\beta P^4$$

where $F(P)$ is free energy, $P$ is polarization and paraelectric-ferroelectric transition occurs when $\alpha = 0$.

In contrast, for the improper ferroelectric YMnO$_3$, where the polarization arises due to a nontrivial coupling to a zone-boundary lattice instability, the $K$ mode, a simplified free energy is given by[71]:

$$\mathcal{F}(P, K) = \alpha_{02} P^2 + \alpha_{20} K^2 + \beta_{40} K^4 + \beta_{31} K^3 P + \cdots$$

where $K$ corresponds to the primary order parameter associated with the zone-boundary $K_3$ phonon mode and $P$ is the polarization. Here the quadratic coefficient of the polarization, $\alpha_{02}$, does not soften to zero at the ferroelectric transition[71, 72], i.e., $\alpha_{02} > 0$ at any temperature. Instead the spontaneous polarization arises because of the $\beta_{31}$ coupling, where for small $K$, $P \sim K^3$, and for large $K$, $P \sim K$[72].

Another prominent improper ferroelectric is the SrTiO$_3$/PbTiO$_3$ superlattice[73], where the polarization arises due to the combination of two rotational modes $R_1$ and $R_2$. And the two rotational modes couple linearly to the polarization, that is, there is a trilinear term in the free energy of the form[69, 74]:

$$\mathcal{F}_{111} = \gamma P R_1 R_2$$

where $P$ is polarization. This type of improper ferroelectrics is thus called trilinear coupling ferroelectricity. The term "hybrid improper ferroelectricity" is also used to describe this ferroelectric mechanism in order to generalize the idea to include cases where the two distortion patterns do not necessarily condense at the same temperature[68].

From the above discussion, we can see that improper ferroelectricity arises not because of zone-center soft phonon instability, as in proper ferroelectrics, but because of zone-boundary rotational modes. Since these rotational modes are due to the combinations of the different radii of the oxygen, A-site cation and B-site cation in ABO$_3$ perovskites, improper ferroelectricity is said to have a (non-electrostatic) geometric-lattice mechanism and hence it is believed to be resistant to electrostatic screening by itinerant charge carriers[69]. In addition, because of the geometric-lattice mechanism, the polarization of improper ferroelectrics is also found to be resistant to the electrostatic forces provided by depolarization field, i.e. the spontaneous polarization of ultrathin improper ferroelectrics is stable in the absence of metallic electrodes[71]. By selecting the suitable chemical species, one can properly design the geometric factor and realize rotation induced polarization in perovskite oxides[69].

After explaining hybrid improper ferroelectricity, let us now discuss the mechanism of the polar metal Ca$_3$Ru$_2$O$_7$[44]. Ca$_3$Ru$_2$O$_7$ belongs to the Ruddlesden−Popper oxides with a general formula of A$_{n+1}$B$_n$O$_{3n+1}$. Ca$_3$Ru$_2$O$_7$ consists of CaRuO$_3$ perovskite blocks stacked along the [001] direction with an extra CaO sheet inserted every 2 perovskite unit cells, taking a group symmetry $Bb2_1m$[44, 68]. The structure exhibits RuO$_6$ octahedral rotations and tilts about [001] and [110], respectively[44]. These main lattice modes can be identified as a polar zone-center mode $\Gamma_5^-$, and two zone boundary modes at the X (1/2, 1/2, 0) point—an oxygen octahedron rotation mode $X_2^+$ and an oxygen octahedron tilt mode $X_3^-$, as shown in Fig. 2[44, 68]. It was found that the polar distortion in Ca$_3$Ru$_2$O$_7$ is mainly due to Ca and O displacement, which arises from a trilinear anharmonic interaction of the form $\alpha Q(\Gamma_5^-)Q(X_2^+)Q(X_3^-)$[44, 68]. Here $Q$ stands for the amplitude of the mode and $\alpha$ is a constant. One can see this mechanism is similar to the trilinear coupling or hybrid improper ferroelectricity. The Ca-O displacement is not screened because the Fermi surface originates from Ru$^{4+}$ orbitals and Ru contributes little to the $\Gamma_5^-$ mode, while the rotational mode $X_2^+$ and the tilt mode $X_3^-$ have geometric origins and hence are negligibly affected by the itinerant charge carriers[44, 68].

One can clearly see that both the lone-pair mechanism and the "hybrid improper ferroelectricity" mechanism fit neatly into the DEM model in that the inversion symmetry is broken by the A-site atom while the conductivity is mainly contributed by the B-site atom.

### 4.2 LiOsO$_3$

As the first unambiguous polar metal, the discovery of LiOsO$_3$ in 2013 has attracted significant research interests[15, 17, 41, 42, 75-80]. At room temperature, LiOsO$_3$ takes a centrosymmetric structure ($R\bar{3}c$), which can be viewed as a result of the $a^-a^-a^-$ octahedral tilting from the higher-symmetry cubic phase ($Pm\bar{3}m$). A phase transition from the centrosymmetric $R\bar{3}c$ structure to the polar $R3c$ structure takes place around $T_s \approx 140$ K accompanied by a shift of about 0.5 Å in the mean positions of the Li atoms along the cubic perovskite [111] axis[15]. The loss of inversion symmetry below $T_s$ is confirmed by neutron diffraction and convergent-beam electron diffraction (CBED) measurements (Fig. 3a-d). In the meanwhile, anomalies in temperature dependence of heat capacity, susceptibility, and resistivity was also observed (Fig. 3e-g)[15]. This phase transition of $LiOsO_3$ is similar to that of $LiNbO_3$ despite the fact that $LiNbO_3$ is an insulator while $LiOsO_3$ is a conductor[81-84].

To form a macroscopic metallic polar phase, the following questions have to be answered: What is the electronic structure and origin of the metallicity of $LiOsO_3$? What is the origin of the polar instability: is it displacive or order-disorder type? What are the driving forces of the long-range ordering of the local dipoles? It is generally accepted that the Fermi level is mainly contributed by the Os and O $d$-$p$ hybridization with minimum contribution from Li[75-77]. The three electrons of $Os^{5+}$ populates the $t_{2g}$ orbitals while $e_g$ orbitals are empty[75].

Over the years, there have been some debates over the origin of the polar instability. Some groups argue it is displacive[17, 41, 42, 75], while others believe it belongs to the order-disorder type[16, 76, 77, 80,]. For the displacive type, the atoms remain associated with their average positions, and phase transition occurs as the Li atoms move along the polar axis and changes its symmetry (Fig. 4a,b). For the order-disorder type, the structural model involves partially occupied sites, and the transition occurs as the symmetry of the occupational distribution is broken (Fig. 4c,d). If the corresponding phonon mode frequency decreases to zero near $T_s$, it is a signature of a displacive case (soft mode behavior). In the order-disorder case, the relevant phonon frequency stays temperature independent[85].

We first discuss the views of the pro-displacive group. Through phonon mode analysis, Xiang found that the polar instability of LiOsO$_3$ can be explained by the softening of the zone center $A_{2u}$ (or $\Gamma_{2-}$) phonon mode, which is mainly dominated by the displacement of the Li atom with a small contribution from the O atoms (Fig. 4a,b)[41]. This picture is well supported by other theoretic works from refs. 17 and 75[17, 75]. Although it is generally accepted that Li displacements play a major role in the polar instability of LiOsO$_3$, there is some dispute over the contribution from Os atoms. Xiang argues that the 5d $t_{2g}$ states of the Os$^{5+}$ ion are partially occupied, and hence the second-order Jahn-Teller effect (or the *d-p* hybridization) could not take place[41]. Whereas Giovannetti and Capone contends that although the 5*d* $t_{2g}$ orbitals are half-filled, the $e_g$ orbitals are empty and open for *d-p* hybridization, which enhances polar distortions, as in common ferroelectric insulators, such as LiNbO$_3$[75].

We next discuss the polar instability from the perspective of the order-disorder mechanism. Based on the fact that the 140 K (~ 12 meV in energy term) phase transition temperature is much smaller than the depth of the double potential wells, which are around 44 meV, Liu et al. argued that the phase transition is of order-disorder type (Fig. 4c,d)[77, 86]. This assertion is supported by the experimentally observed incoherent charge transport in LiOsO$_3$ above the transition temperature (Fig 3g), which could be attributed to the scattering induced by disorder of Li off-center displacement[15, 77]. In addition, it is further supported by the absence of the $A_{2u}$ soft phonon mode in the Raman spectroscopy measurements by Jin et al.[80]. However, the $A_{2u}$ soft phonon is recently detected by Laurita by ultrafast optical pump–probe experiments[42].

These contradicting arguments over the displacive or order-disorder nature of the polar instability may suggest that LiOsO$_3$ has characters of both, as suggested by Sim et al.[76] and Laurita et al.[42]. This fact should be understandable as even the prototype "displacive" ferroelectric BaTiO$_3$ was found to have both displacive and order-disorder components[13, 85, 87, 88]. In addition, LiNbO$_3$ has also been suggested to have both displacive and order-disorder characters[89]. It may be worthwhile to carry out some nuclear magnetic resonance (NMR) experiments to clarify the origin of polarity in LiOsO$_3$ as done in BTO[87].

Next, we review the theories proposed to explain the driving forces of the long-range ordering of the local dipoles. The key here is that if there is no interaction between these local dipoles, then the local dipoles may be disordered and the system will be non-polar macroscopically. By employing an effective Hamiltonian to study the interaction between local polar modes, Xiang found that the short-range interactions between some local dipole modes survive the electronic screening and stabilizes the ferroelectric-like ordering (Fig. 5a,b)[41]. On the other hand, by approximating the dipole interactions with Li-Li pairs, Liu et al. found that these interactions are only slightly screened along certain directions (for example, there is almost no conduction charge distribution between pair 1 in Fig. 5c) and hence long-range ordering is possible along these directions[73]. This is because that the electronic density mainly concentrates between Os and O ions due to the strong hybridization between Os 5$d$ and O 2$p$, while there is almost no conduction charge in a relative large space around Li ion (Fig. 5c-e)[77]. This incomplete screening scenario is consistent with the claim by Vecchio et al.[79], which states that LiOsO$_3$ is a strongly correlated bad metal close to Mott localization due to the half-filling configuration of the Os 5$d$ $t_{2g}$ orbitals. It is also supported by the fact that the resistivity of LiOsO$_3$ exceeds that of a normal metal by two orders of magnitude[24] since higher resistivity means less effective electronic screening.

These theoretical works strongly supports the DEM model proposed by Puggioni and Rondinelli[30]. It is recently proved in the ultrafast spectroscopy experiments by Laurita et al., in which they demonstrated that the intra-band photo-carriers relax by selectively coupling with a subset of the phonon spectrum ($^1E_g$ and $^2E_g$), while they couple extremely weakly to the $A_{2u}$ soft polar mode[42]. In addition, this DEM model is also supported by the dual nature of the $d$ orbitals of Os atoms, in which the electrons in $t_{2g}$ orbitals are responsible for the conductivity while the empty $e_g$ orbitals hybrid with O $2p$ orbital to enhance the polar distortions as in common ferroelectric insulators[75]. While the origin of the polar instability and the mechanism of the long-range ordering of local dipoles in LiOsO$_3$ have been well explained, its switchablity by external electric field has not been demonstrated experimentally so far, despite theoretical predictions[41, 90]. In the next section, we review the progress in the first electric-field-switchable ferroelectric metal WTe$_2$.

**4.3 WTe$_2$**

As a member of the transition metal dichalcogenides (TMDCs) family, WTe$_2$ has been extensively studied over the past years due to its exotic topological properties[91-94]. It takes a layered orthorhombic structure (1T′), which belongs to the polar space group $Pmn2_1$[95, 96]. While two-dimensional ferroelectricity has been widely reported in a variety of two-dimensional materials, such as SnTe[97], CuInP$_2$S$_6$[98] and IV-VI group compound[99], most of them are semiconductors. Interestingly, WTe$_2$ is a Weyl semimetal[91]. In 2018, through detailed electrical transport measurements under external electric field by sandwiching 2-3 WTe$_2$ layers between two hexagonal boron nitride dielectric sheets, Fei et al. presented WTe$_2$ as the first external-electric-field-switchable ferroelectric metal with an out-of-plane polarization (Fig. 6a,b)[19]. Later, Sharma et al. demonstrated that WTe$_2$ is a ferroelectric metal even in its bulk form through extensive piezoresponse force microscopy (PFM) measurements (Fig. 6c-g)[20].

The natural questions following the demonstration of the first ferroelectric metal are: what is the origin of the ferroelectricity in WTe$_2$ and what is its switching mechanism? Following the suggestion by Fei et al. that "the polarization could principally involve a relative motion of the electron cloud relative to the ion cores, rather than a lattice distortion", Yang et al.[100] and Liu et al.[101] demonstrated that the polarization stems from uncompenstated interlayer charge transfer. The calculated charge transfer of ~ $3.2 \times 10^{11}$ e/cm$^2$ is consistent with the experimentally measured polarization of $2 \times 10^{11}$ e/cm$^2$ at 20 K[19]. This charge transfer can be clearly seen in the differential charge diagram shown in Fig. 7 between the upper and lower layers of WTe$_2$. These authors continue to suggest that the out-of-plane polarization can be switched by an in-plane interlayer sliding[100, 101], which has been used to explain the switchability of other two-dimensional van der Waals materials, such as In$_2$Se$_3$[102, 103] and MoS$_2$[104]. From Fig. 7a, one can see that the polarization can be switched by moving the upper layer along the −y axis by a distance of $\Delta_1 + \Delta_2$, as evidenced by the inversion of the charge distribution between the upper and lower layers in Fig. 7b-c. Yang et al. also showed that the switchability of multilayer WTe$_2$ can also be described by this model by selectively sliding some specific layers[100].

This interlayer sliding model explains some important aspects in WTe$_2$, but also left some critical questions unanswered. For example, how does the out-of-plane electric field induce an in-plane interlayer sliding? In addition, the magnitude of the required interlayer sliding ($\Delta_1 + \Delta_2$ ~ 50 pm) is much larger than the atomic distortions (~ 10 pm) in conventional ferroelectric switching[103]. It can be experimentally tested with microscopic structural characterization tools such as STEM. This model also imposes some limitations. For example, it needs to selectively slide the layers to switch the polarization of multilayers[100], which is unrealistic to explain the switchability of bulk WTe$_2$.

The sliding mechanism of polarization switching in WTe$_2$ is reminiscent of the decoupled electron mechanism (DEM) model in that the orientation of polarization and the direction of the polarization switching are spatially separately, which we would like to call "decoupled space mechanism (DSM)". We also note that another paper has proposed a similar switching mechanism of ferroelectric metals through interfacial coupling with a ferroelectric insulator, where the polarization and its switching are located in different regions in the system. Fang et al. proposed BiPbTi$_2$O$_6$ to be a polar metal, where the conductivity comes from the Ti $3d$ states and the polar instability originates from the $6s$ lone-pair electrons in Bi and Pb similar to BiFeO$_3$ and PbTiO$_3$[45]. By growing BiPbTi$_2$O$_6$ on PbTiO$_3$ with in-plane ferroelectricity, which can be achieved with tensile strain[105, 106], the in-plane polarization of BiPbTi$_2$O$_6$ can be switched if that of PbTiO$_3$ is switched[45]. This approach circumvents the need to directly apply an external voltage on the polar metal, which will induce an electric current instead of an electric field. We hope this DSM mechanism may give researchers some hints to study the large number of polar metals whose swichability is yet to be demonstrated. Although some important progress has been made in understanding the physics of the first ferroelectric metal WTe$_2$, some important questions need to be investigated. For example, is the ferroelectricity a type of prototypical proper ferroelectricity, hyperferroelectricity (discussed below) or improper ferroelectricity? Can the ferroelectric metallic state be explained by the DEM model?

**4.4 Hyperferroelectric metals**

In the above section discussing improper ferroelectrics, we mentioned that because of its geometric-lattice origin of polarization, it is resistant to the electrostatic forces provided by itinerant charge carriers and depolarization field. In 2014, Garrity, Rabe, and Vanderbilt proposed another type of ferroelectricity, coined hyperferroelectricity, which is also resistant to depolarization field[107]. Since the polarization of hyperferroelectric is resistant to the electrostatic forces provided by depolarization field, one natural question we can ask is whether it can persist under the electrostatic forces provided by itinerant charge carriers. Recently, several hyperferroelectric metals have been proposed, such as in LaPtSb[53], LaAuGe[53], SrHgPb[49], KMgSb$_{0.2}$Bi$_{0.8}$[50], CaAgBi[51], CrN[55] and CrB$_2$[55].

Let us briefly introduce hyperferroelectricity and discuss its differences with prototypical proper ferroelectricity and improper ferroelectricity before summarizing the recent progress in hyperferroelectric metals. Hyperferroelectrics is first proposed by Garrity, Rabe, and Vanderbilt in hexagonal ABC compounds (LiGaGe type)[107]. Fig. 8a and 8b show the non-polar high-symmetry $P6_3/mmc$ structure and the polar $P6_3mc$ structure, respectively[108]. The structure is a hexagonal variant of the half-Heusler structure and can be described as a wurtzite structure ''stuffed''with a third cation (the A atom in ABC notation)[108]. The polar phase is reached primarily by a buckling in the honeycomb layers as the atoms move from a $sp^2$ environment towards $sp^3$ bonding, resulting in polarization in the z direction[107, 108].

Let us first discuss the difference between hyperferroelectrcity and prototypical proper ferroelectricity. Hyperferroelectricity is a type of proper ferroelectricity, in this article we use the term "prototypical proper ferroelectrics" to describe those proper ferroelectrics other than hyperferroelectrics. The ferroelectric phase transitions can be understood by the lattice dynamics of their high-symmetry phase. For proper ferroelectrics, the high-symmetry phase has at least one unstable TO mode, specifically, a $\Gamma$ mode that is unstable under zero macroscopic electric field ($\varepsilon = 0$) boundary conditions[107]. The frequency of this mode can be obtained from first-principles computation of the force-constant matrix with the usual periodic boundary conditions[107]. If the depolarization field is unscreened, corresponding to the case of electric displacement $D = 0$, the structure instability is determined by the LO modes, which can be obtained by adding to the dynamic matrix a non-analytic long-range Coulomb term that schematically takes the form $(Z^*)^2/\varepsilon^\infty$, where $Z^*$ is the Born effective charge and $\varepsilon^\infty$ is the electronic contribution to the dielectric constant, generating the well-known LO-TO splitting[107-110]. In prototypical proper ferroelectrics, such as BTO and PTO, due to their large Born effective charges and relatively small $\varepsilon^\infty$, the LO-TO splittings are huge, such that all LO modes are stable[107-110]. Therefore they lose ferroelectricity if the depolarization field is not well screened. In contrast, in the hexagonal ABC hyperferroelectrics, the LO-TO splittings are small, such that even the LO modes can become unstable[107-110]. The unusual electric properties of hyperferroelectrics mean that the ferroelectric phase transition temperature, which decreases with decreasing screening of the depolarization field, will still be nonzero even under $D = 0$ boundary conditions[107-110]. At this temperature $T_D$, the LO mode becomes unstable and the material becomes a hyperferroelectric[107-110]. Consequently, the polarization in these materials is very robust against the depolarization field. Although hyperferroelectrics share the same depolarization-field-resistant property as improper ferroelectrics, it also differs significantly with improper ferroelectrics in the primary order parameters and the phonon modes. As hyperferrroelectrics is a type of proper ferroelectrics, the differences between proper and improper ferroelectics discussed in section 4.1 also applies here, interested readers can find more comprehensive discussions in refs. 107-110.

After introducing hyperferroelectricity, let us now discuss hyperferroelectric metals introduced at the beginning of this section. Strictly speaking, the term "hyperferroelectric metal" or even "hyperferroelectricity" is not scientifically rigid at this moment because their polarization has not yet been experimentally shown to be switchable, to the best of our knowledge. But as discussed in section 2, the misuse of terms has happened a lot in this confusing field that studies the coexistence of contra-intuitive properties. A more proper term for these materials would be "hyperpolar metals" before its switchability is shown. Anyhow, we decide to follow the convention and continue to call them hyperferroelectric metals in this work.

Let us take LaPtSb as an example of hyperferroelectric metal[53]. It takes the typical hexagonal ABC structure as shown in Fig. 8. Recently, Du et al. found that LaPtSb is a hyperferroelectric metal with a resistivity that is nearly an order of magnitude lower than the well-studied oxide polar metals[53]. By utilizing a Density Functional Theory (DFT) Chemical Pressure (CP) analysis to analyze the local bonding interactions between La, Pt and Sb, they found that the bucking of the Pt-Sb (or B-C in ABC notation) plane is a result of the relief of intraplanar positive chemical pressures created by local bonding preferences[53]. Because these bonding preferences are local, the resulting polar bucking is reasonably expected to be resistant to electrostatic screening by itinerant charge carriers. Despite some progress has been made in hyperferroelectric metals, there have been some major issues left unsolved. Such as the experimental observation of the LO soft phonon and of course the switchability of polarization.

## 5. Design principles for ferroelectric/ polar metals

After reviewing the major categories of ferroelectric/polar metals, we believe it is appropriate to summarize the routes or design approaches for ferroelectric/ polar metals. Of course one can try to sort out all polar metals by scrutinizing material libraries, such as the Inorganic Crystal Structure Database (ICSD), as done by Benedek and Birol in ref. 17. But this method is time-consuming and the libraries cannot include new materials which has yet to be discovered. If one wants to artificially design a polar/ferroelectric metal, it would be helpful to have some guiding principles. Ultimately, the polar/ferroelectric metal problem is a structure-property problem. So we can view this problem from two perspectives. From the perspective of structure, the polar/ferroelectric metal problem is equivalent to the preservation of polar structures under the influence of itinerant charge carriers. For those materials which take the perovskite $ABO_3$ structure or the layered perovskite structure, it is important to carefully design the geometric or chemical factor to let the polar distortion originates mainly from the A-O displacement, such as PTO[35], BFO[37], LiOsO$_3$[15], NdNiO$_3$[40], (Sr,Ca)Ru$_2$O$_6$[30] and Ca$_3$Ru$_2$O$_7$[44] . Because the conductivity of these materials almost always comes from the orbital hybridization between B atom and oxygen. Thus, based on the DEM model, the electronic degree of freedom is decoupled from the inversion-symmetry-breaking degree of freedom[30]. Although it looks like the DEM model explains the origin of most oxide-based polar metals, it should be noted that there are also some exceptions[17]. For example, in the hypothetical material TiGaO$_3$, the Ti atom is responsible for both polar instability and metallicity[41]. Moreover, it remains whether the LiGaGe-type and other two-dimensional ferroelectric/polar metals (e.g. WTe$_2$) can be explained by the DEM model. In addition, although the perovskite ABO$_3$ materials with B-type displacement is prone to electrostatic screening by itinerant charge carriers, researchers have found that the polar distortion can be sustained by the "meta-screening effect"[39] and enhanced by strain[62].

From the perspective of ferroelectricity, we have seen that hyperferroelectrics and improper ferroelectrics are pretty immune to electrostatic forces provided by depolarization field and itinerant charge carriers. Thus these materials provide fertile ground for studying polar/ferroelectric metals. One potential route for novel polar/ferroelectric metals is to find conducting hyperferroelectrics and improper ferroelectrics. Another potential route involves charge carrier doping into insulating hyperferroelectrics and improper ferroelectrics.

## 6. Future prospects and summary

In this article, we reviewed the history of ferroelectric/polar metals, the various milestones in this field and the mechanisms of the ferroelectric/polar metallic phase in these materials. Finally, we summarized the design principles for ferroelectric/polar metals. Here we would like to briefly discuss the future prospects.

Although many polar metals have been discovered over the past years, only WTe$_2$ has been demonstrated to be switchable. Despite that an external electric field is expected to induce an electric current in these materials and make switching difficult, several switching mechanisms have been proposed and deserve more experimental research. For example, a thin enough polar metal may be sufficiently penetrated by an external electric field to have its polarity switched as in WTe$_2$[19]. In addition, materials with limited reservoir of mobile charges may lead to incomplete screening in the polar metal and offers the potential of switchability as proposed in Bi$_5$Ti$_5$O$_{17}$[31]. Moreover, the DSM model has been demonstrated to be able to switch the polarization of BiPbTi$_2$O$_6$ by indirectly coupling to a ferroelectric insulator as proposed in ref. 45. Furthermore, as suggest by Wang et al. in ref. 12, polarity switching is possible in polar metals if the applied voltage rises fast enough in time[111]. Finally, there exist means to switch polarization without applied voltage, such as mechanical switching[112] and chemical switching[113].

With the discovery of the coexistence of ferroelectricity/polarity and conductivity in a single phase, a lot of intriguing properties have been reported in ferroelectric/polar metals, such as unconventional superconductivity[58, 114, 115], unconventional optical responses[116, 117], magnetoelectricity[57, 118, 119] and thermoelectricity[30]. Based on these properties, some prototypical applications have been theoretically proposed. Puggioni and Rondinelli proposed an (Sr,Ca)Ru$_2$O$_6$-based anisotropic thermoelectric device based on the fact that it exhibits anisotropic Seebeck coefficients derived from the polar structure (Fig. 9a)[30]. The peculiar thermopower anisotropy dictates that the electric field resulting from an applied heat flux to the material will be non-collinear[30, 120]. It enables the heat flux to be measured in a geometry perpendicular to the induced electrical current, specifically at locations where the temperatures are equal[30, 120]. In addition, because of the metallic nature of (Sr,Ca)Ru$_2$O$_6$, the relaxation time is small compared to conventional semiconductor- or insulator-based thermoelectric devices. This enables $(Sr,Ca)Ru_2O_6$-based thermoelectric device to find applications in ultrafast measurements[30, 121]. The same group also proposed another application where the polar metal is used as electrodes to suppress the critical-thickness limit, which disables us from continuous scaling changes demanded by higher density data storage technologies, in ferroelectric nanocapacitors[122]. Fig. 9b shows the structure of a [LiO-(OsO$_2$-LiO)$_n$/NbO$_2$-(NaO-NbO$_2$)$_m$] ($m$=1, $n$=6) ferroelectric nanocapacitor. One can see that the polarization of the relaxed structure is preserved even when $m$=1. The persistence of the polarization of NaNbO$_3$ is due to the interfacial coupling between the polar displacements of NaNbO$_3$ and LiOsO$_3$. This effect does not rely on interfacial bond chemistry or "perfect" screening of the depolarizating field, but rather results from the intrinsic broken parity present in the LiOsO$_3$ electrode. Based on the finding that the in-plane spin polarization is reversed upon vertical ferroelectric switching of bilayer WTe$_2$, Liu et al. proposed a spin field electric transistor (spin-FET) as shown in Fig. 9c[101]. The electrodes are composed of two multilayer ($\geqslant$ 3) WTe$_2$ covered with a ferromagnetic metal, the middle channel structure is composed of a bilayer WTe$_2$, and the gate and back electrodes allow the application of the required voltage to switch the dielectric polarization of the bilayer WTe$_2$. Based on the model that the polarization and hence the electric potential between bilayer WTe$_2$ can be switched by interlayer sliding, Yang et al. proposed a bilayer WTe$_2$-based nanogenerator for harvesting energy from human activities, ocean waves, mechanical vibration, etc[100, 123]. It should be noted, however, despite these fascinating proposals, most of them are hypothetical at the moment and deserve more experimental evidence for future applications in fields of photovoltaics, electronics and spintronics. We hope our work will help researchers understand the progress in ferroelectric/polar metals and encourage further investigations into this fascinating field.

**Figures**

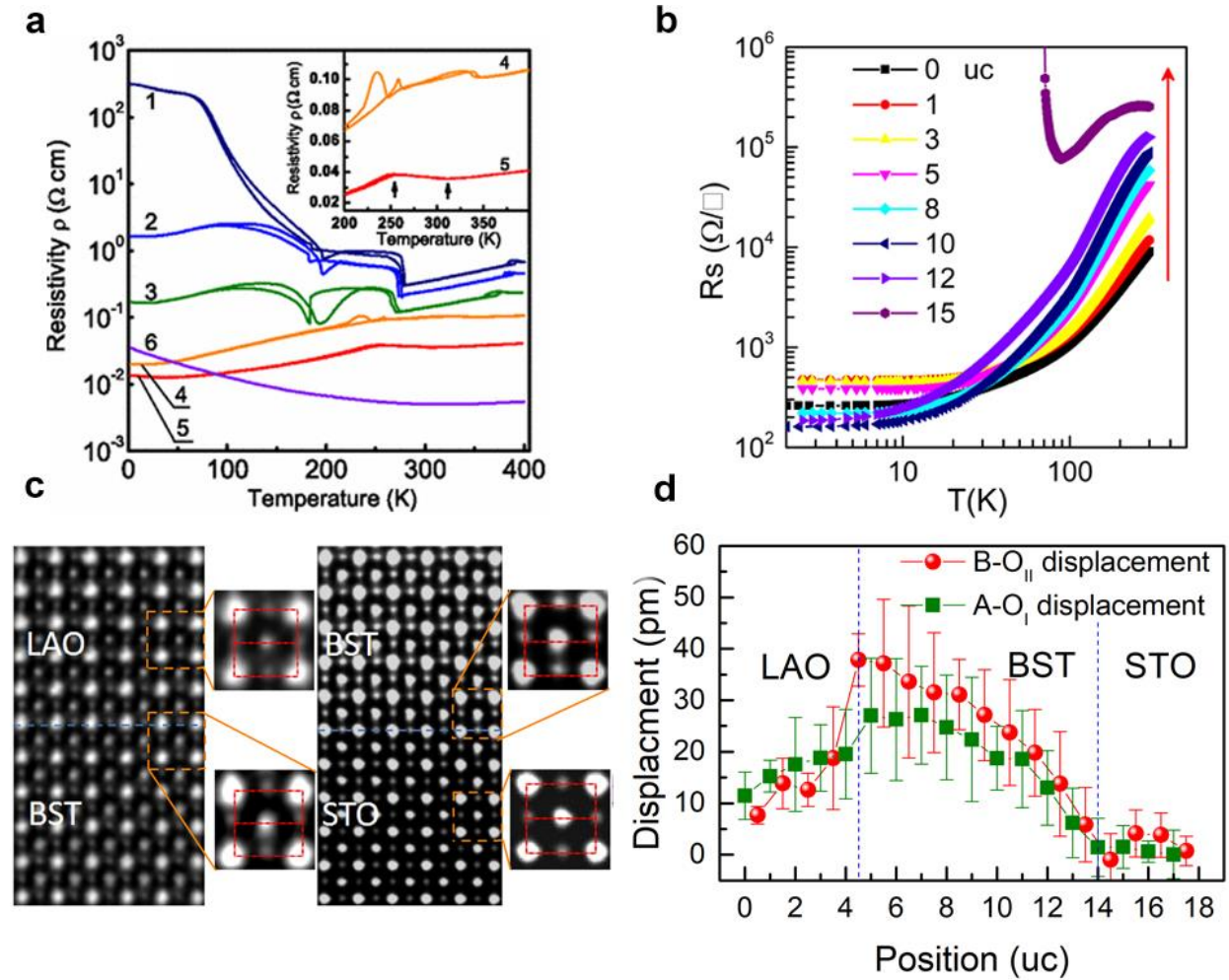

**Fig. 1.** Electronic transport properties of bulk $BaTiO_{3-\delta}$ and the polar metallic phase in $Ba_{0.2}Sr_{0.8}TiO_3$ thin films. (a) Temperature dependent resistivity $\rho(T)$ of $BaTiO_{3-\delta}$ measured on both cooling and heating to reveal temperature hysteresis of the phase transitions. Electron concentration $n$ was determined from the Hall effect. Samples 1, 2, 3, 4, 5 correspond to $n = 1.9 \times 10^{19}$, $3.1 \times 10^{19}$, $6.1 \times 10^{19}$, $1.6 \times 10^{20}$, and $3.5 \times 10^{20}$ cm$^{-3}$, respectively. Sample 6 is a $BaTi_{0.875}Nb_{0.125}O_3$ polycrystal sample with $n \approx 2.0 \times 10^{21}$ cm$^{-3}$. The $\rho(T)$ dependence changes from insulating (curves 1 and 2) to metallic (curves 4 and 5). Structural phase transitions are manifested by pronounced $\rho(T)$ anomalies at corresponding temperatures. The inset shows an enlarged part of resistivity for metallic samples 4 and 5 in the region of the orthorhombic-tetragonal (O-T) and tetragonal-cubic (T-C) phase transitions. The O-T and T-C phase transitions for sample 5 are indicated by arrows. The disappearance of $\rho(T)$ anomaly in sample 6 suggest that at $n > n_c \approx 1.9 \times 10^{21}$ cm$^{-3}$ the polar ground state in metallic $BaTiO_3$ is destroyed. Adapted with permission from ref. 27. DOI: https://doi.org/10.1103/PhysRevLett.104.147602. Copyrighted by the American Physical Society[27]. (b) Temperature dependent sheet resistance $R_s$ of $LaAlO_3/Ba_{0.2}Sr_{0.8}TiO_3/SrTiO_3$ heterostructures with fixed $LaAlO_3$ thickness (15 unit cells) and different unit cells of $Ba_{0.2}Sr_{0.8}TiO_3$ on $SrTiO_3$ substrates. Different colors represent different $Ba_{0.2}Sr_{0.8}TiO_3$ thicknesses. The red arrows indicate increasing BST thickness. All the samples

except the one with 15 unit cells of $Ba_{0.2}Sr_{0.8}TiO_3$ show fully metallic behavior from 300 to 2 K. (c) Atomically resolved inverted annular-bright-field scanning transmission electron microscopy (ABF-STEM) images of the $LaAlO_3$/ $Ba_{0.2}Sr_{0.8}TiO_3$ interface and $Ba_{0.2}Sr_{0.8}TiO_3$/$SrTiO_3$ interface of a $LaAlO_3$/$Ba_{0.2}Sr_{0.8}TiO_3$/$SrTiO_3$ heterostructure with 15 unit cells of $LaAlO_3$ and 10 unit cells of $Ba_{0.2}Sr_{0.8}TiO_3$. The displacements of Ti and O from their equilibrium positions can be clearly seen in $Ba_{0.2}Sr_{0.8}TiO_3$. (d) Out-of-plane $B$-$O_{II}$ and $A$-$O_I$ displacements across the $LaAlO_3$/$Ba_{0.2}Sr_{0.8}TiO_3$/$SrTiO_3$ heterostructure. Adapted with permission from ref. 34. Copyright 2019, Springer Nature[34].

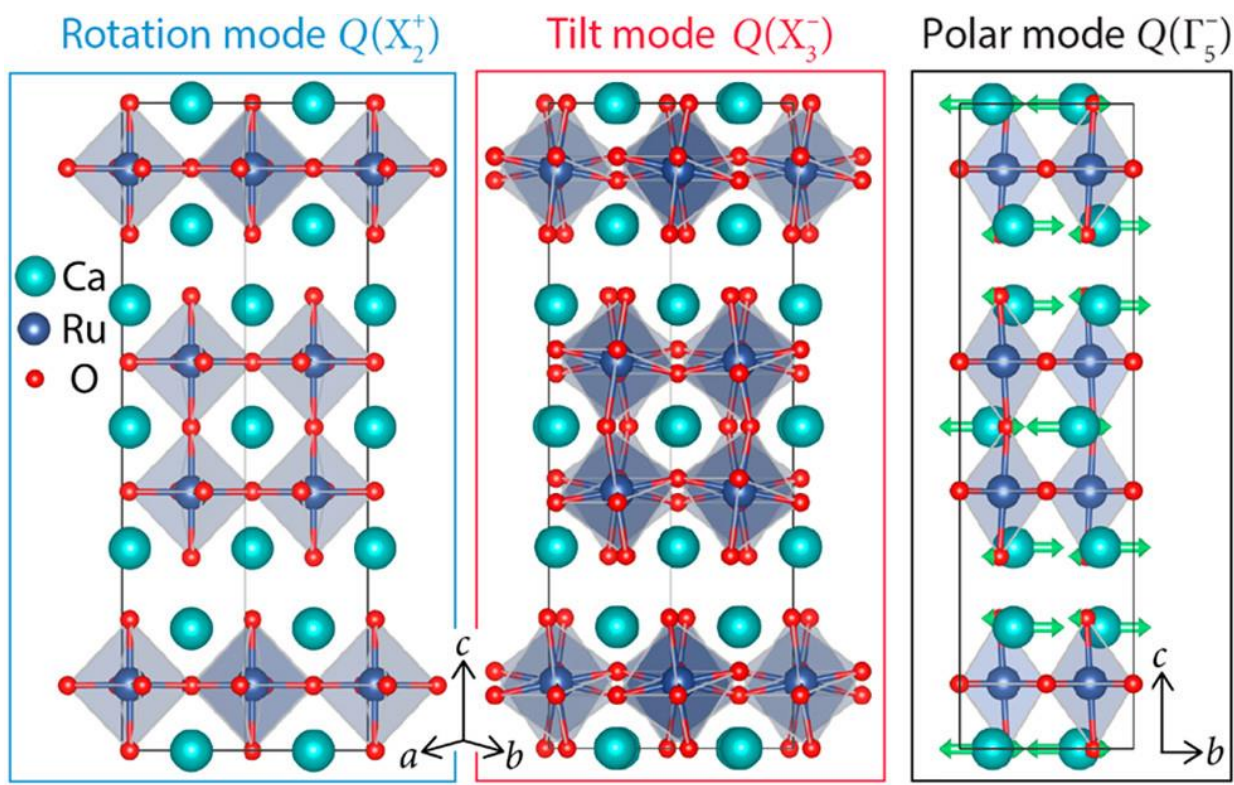

**Fig. 2.** Atomic structures of $Ca_3Ru_2O_7$ and the oxygen octahedron rotation mode $X_2^+$, the oxygen octahedron tilt mode $X_3^-$ and polar mode $\Gamma_5^-$. Adapted with permission from ref. 44. Copyright 2018, American Chemical Society[44)].

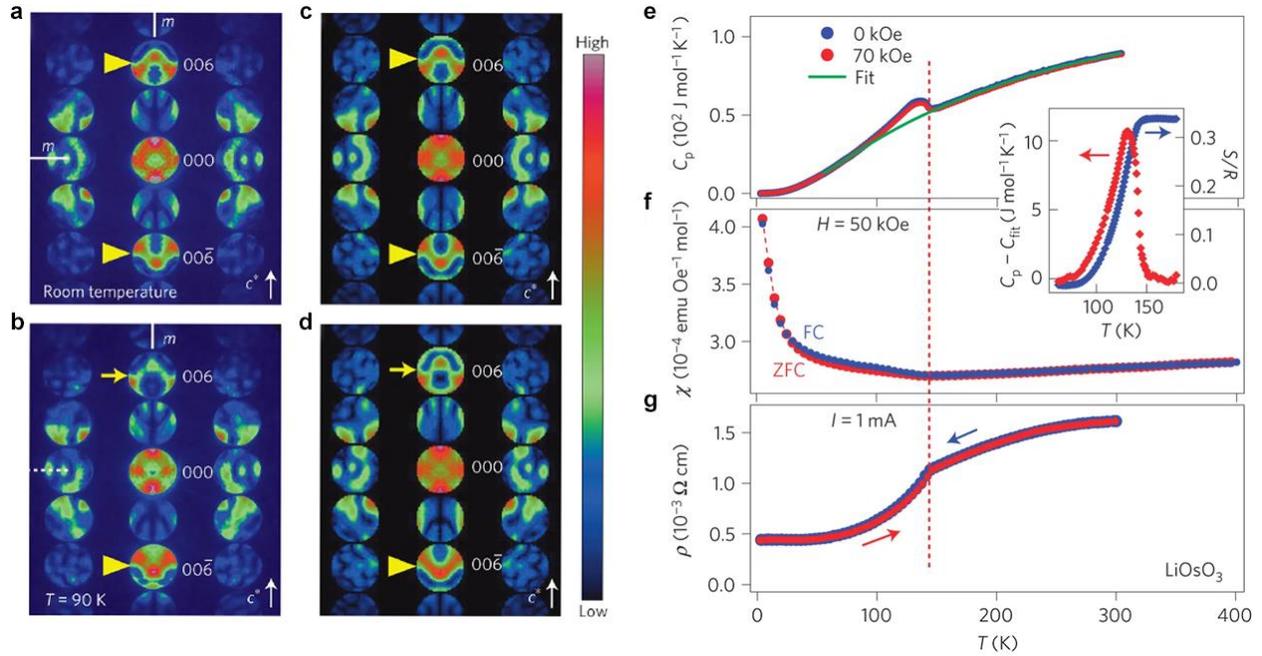

**Fig. 3.** Loss of inversion symmetry below $T_s$ and temperature dependent electrical, magnetic and calorimetric properties of LiOsO$_3$. Convergent-beam electron diffraction (CBED) measurements at room temperature (a) and 90 K (b). Corresponding simulated CBED patterns for a specimen thickness of 73 nm using the centrosymmetric model ($R\bar{3}c$) (c) and the non-centrosymmetric model (*R3c*) (d). An arrow or arrowhead indicates the absence or presence of mirror symmetry perpendicular to the $c^*$ axis. The CBED data clearly shows the room temperature structure is centrosymmetric, while the 90 K structure is non-centrosymmetric. Temperature dependent heat capacity $C_p$ of LiOsO$_3$ (e). The zero-field cooled (ZFC) and field-cooled (FC) magnetic susceptibility $\chi$ of LiOsO$_3$ in a measuring field of 50 kOe (f). Temperature dependent resistivity $\rho$ of LiOsO$_3$ (g). Adapted by permission from Springer Nature Customer Service Centre GmbH: Springer Nature, Nature Materials, ref. 15, Copyright (2013)[15].

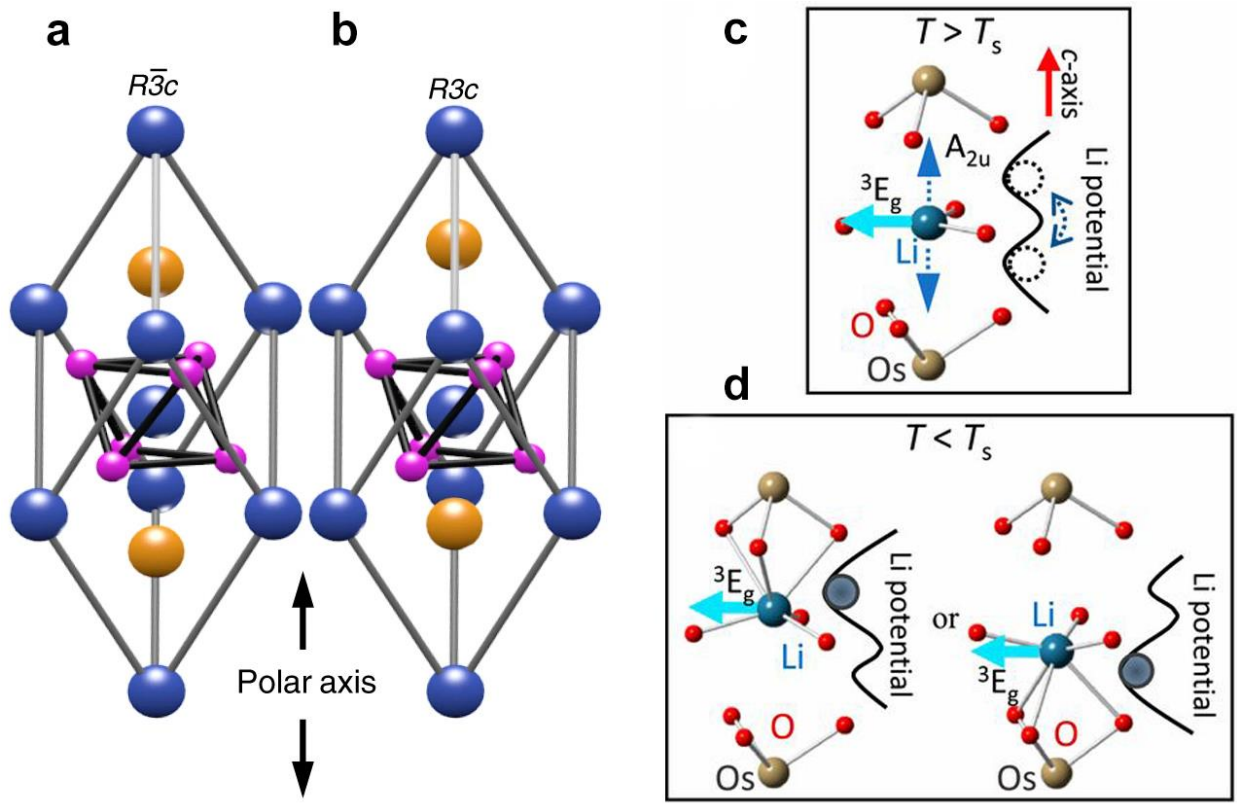

**Fig. 4.** Displacive vs. order-disorder phase transition of LiOsO$_3$. (a) and (b) show the high temperature $R\bar{3}c$ and low temperature $R3c$ structures, respectively, in the displacive scenario of phase transition. Blue, orange, and pink spheres represent Os, Li, and O atoms, respectively. The phase transition is characterized by the displacement of the Li ions along the polar axis (black arrows). Adapted with permission from ref. 42. Copyright 2019, Springer Nature[42]. (c) and (d) show the high temperature $R\bar{3}c$ and low temperature $R3c$ structures, respectively, in the order-disorder scenario of phase transition. Li ions, on average, are located at the center of double wells as the equilibrium position above the transition ($T > T_s$) (c), while the broken centrosymmetry below $T_s$ results in an asymmetric potential well by freezing Li to 1 side of the potential well (d). Adapted with permission from the authors of ref. 16. Copyright 2019 the Author(s). Published by PNAS[16].

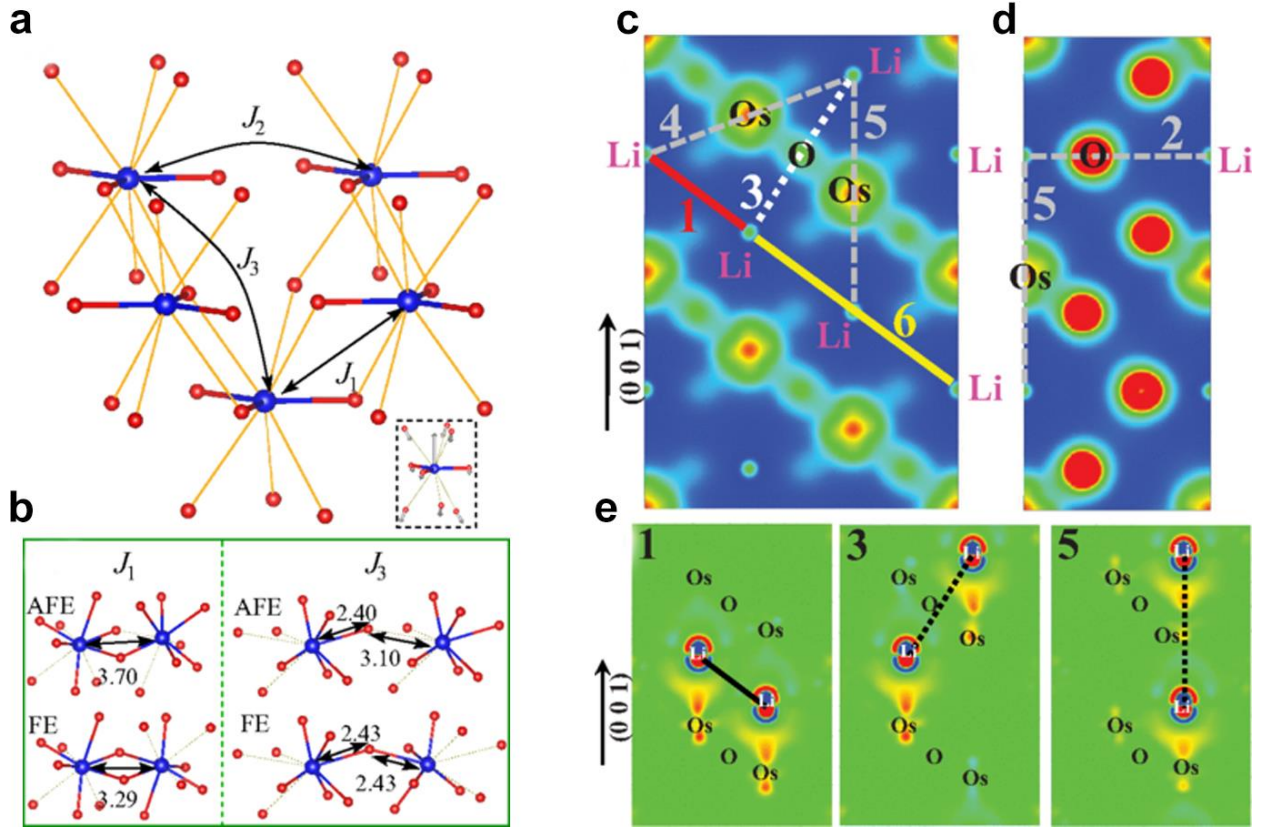

**Fig. 5.** Models of local interactions that stabilize the long-range ordering of dipoles. (a) and (b) show the local modes and the interactions between the local modes. (a) The three interaction paths between the local modes. The displacements in the local mode is displayed in the inset (O displacements are enlarged by ten times for clarity). (b) illustrates why the first ($J_1$) and third ($J_3$) interactions between the local modes are ferroelectric. Red and blue spheres represent oxygen and Li, respectively. Os is omitted for clarity. For detailed discussions on the interactions between the local modes, please refer to ref. 41. Adapted with permission from ref. 41. DOI: https://doi.org/10.1103/PhysRevB.90.094108. Copyrighted by the American Physical Society.[41] Partial electron densities contour maps for paraelectric $LiOsO_3$ taken through [1 -1 0] (c) and [2 -1 0] (d) planes. Contour levels shown are between 0 (blue) and 0.3 e/ Å$^3$ (red). (e) Charge density difference between "ferroelectric" and paraelectric structures for Li pair 1, 3, and 5 through the [1 -1 0] plane. Contour levels shown are between −0.004 (blue) and 0.004 e/ Å$^3$ (red). One can clearly see that dipole interaction in pair 1 is only slightly screened, pair 3 is mildly screened, while pair 5 is heavily screened. Please refer to ref. 77 for detailed discussions. Adapted with permission from ref. 77. DOI: https://doi.org/10.1103/PhysRevB.91.064104. Copyrighted by the American Physical Society[77].

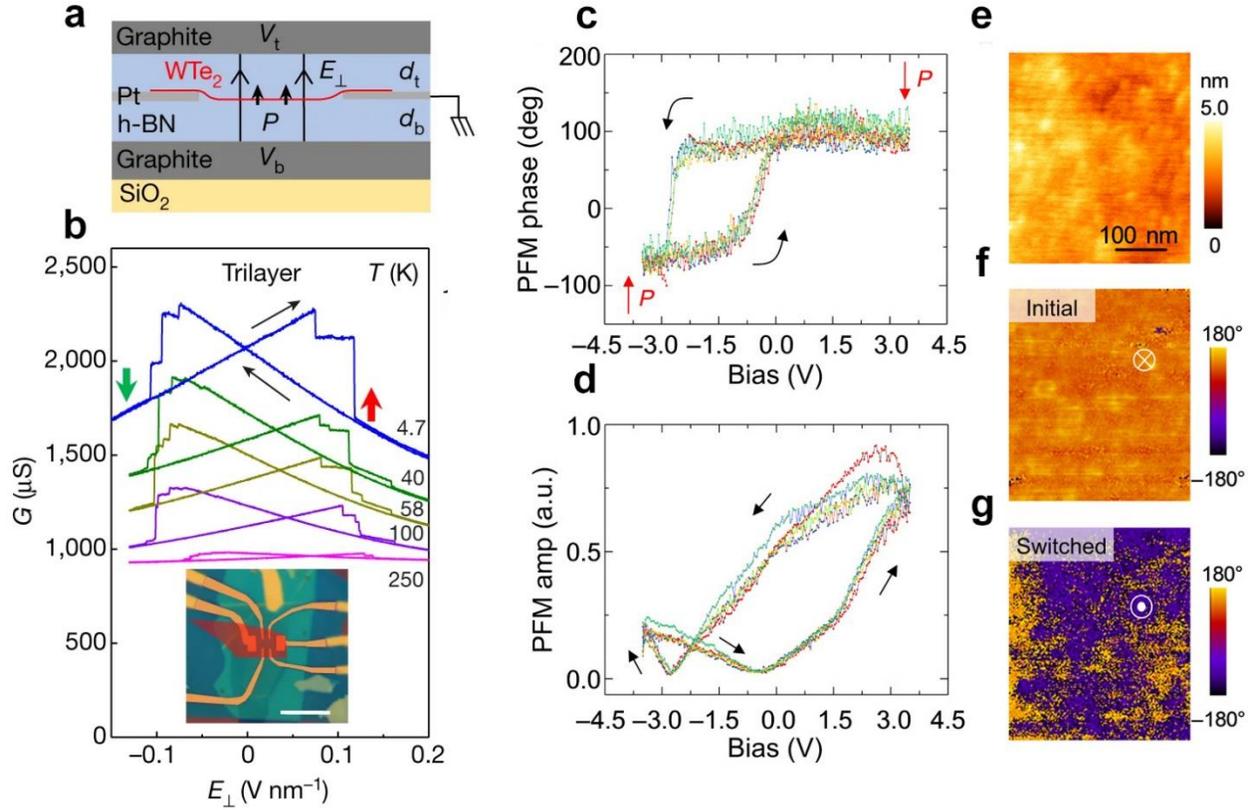

**Fig. 6.** Ferroelectric switching in WTe$_2$. (a) and (b) show electric switching of trilayer WTe$_2$. (a) The capacitor geometry employed to switch WTe$_2$. An electrically contacted thin WTe$_2$ flake is sandwiched between two hexagonal boron nitride (h-BN) dielectric sheets, with thicknesses of $d_t$ (top) and $d_b$ (bottom). Above and below are gate electrodes, usually of few-layer graphene, to which voltages $V_t$ and $V_b$ are applied relative to the grounded WTe$_2$. (b) Conductance $G$ of undoped trilayer device as $E_\perp$ ($E_\perp = (-V_t/d_t + V_b/d_b)/2$) is swept up and down (black arrows), setting $V_t/d_t = -V_b/d_b$ to avoid net doping. The plots show bistability associated with electric polarization up (red arrow) or down (green arrow), at temperatures from 4 K to 300 K (as labelled). Here the conductance is the reciprocal of the four-terminal resistance. The undoped trilayer has a metallic temperature dependence. Inset to (b), optical image of a representative double-gated device. The WTe$_2$ flake has been artificially coloured red. Scale bar, 10 μm. Adapted by permission from Springer Nature Customer Service Centre GmbH: Springer Nature, Nature, ref. 19, Copyright (2018)[19]. (c)-(g) show piezoresponse force microscopy (PFM) measurements of 15 nm WTe$_2$ flake in a capacitor geometry. To see PFM measurements on bulk WTe$_2$, check Fig. 2 of ref. 20. Spectroscopic bias-dependent piezoresponse phase (c) and amplitude (d) hysteretic curves acquired through the top metal electrode gating the WTe$_2$ flake. (e and f) Topography image showing zoom-in on the metal-gated WTe$_2$ (e) and the corresponding piezoresponse phase image (f). (g) PFM phase image after application of a bias pulse of −2.5 V. Adapted by permission from the CC BY-NC 4.0 license. Copyright 2019 the Authors of ref. 20[20].

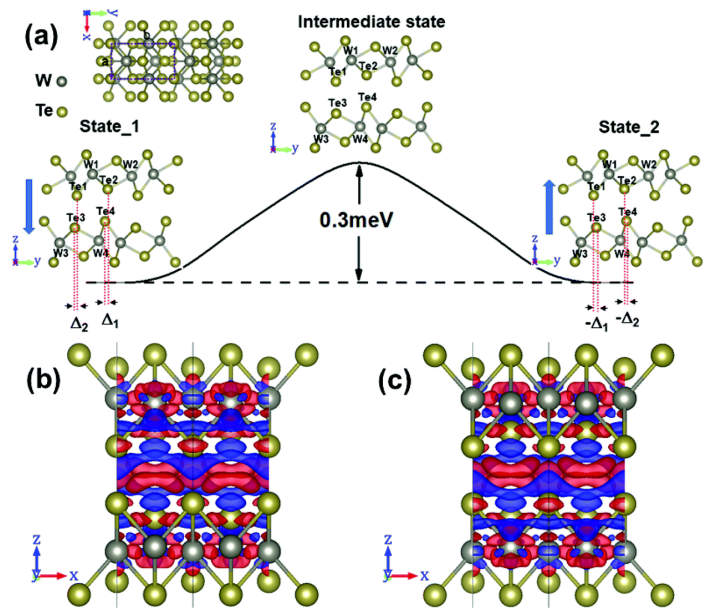

**Fig. 7.** The sliding model for the ferroelectric switching in WTe2. Ferroelectric switching pathway of bilayer WTe2 from bistable State 1 to State 2 (a). Spheres in gray and dark yellow denote W and Te atoms, respectively. The blue arrows indicate the out-of-plane polarization direction. (b) and (c) are the differential charge density diagrams of State 1 and State 2, respectively. The area in red is the region that gained electrons and the area in blue is the region that lost electrons. Republished with permission of Royal Society of Chemistry, from ref. 101; permission conveyed through Copyright Clearance Center, Inc[101].

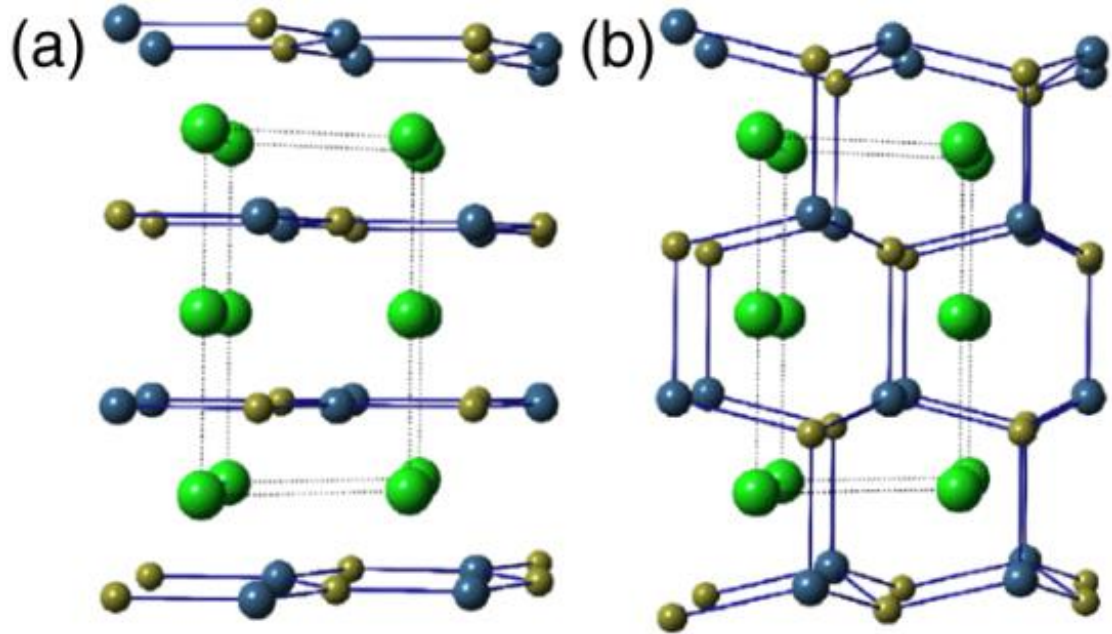

**Fig. 8.** Atomic structures of (a) high-symmetry (P6$_3$/mmc) and (b) polar (P6$_3$mc) ABC ferroelectrics. The large green atom is the stuffing atom (the A atom in ABC notation). Adapted with permission from ref. 107. DOI: https://doi.org/10.1103/PhysRevLett.112.127601. Copyrighted by the American Physical Society[107].

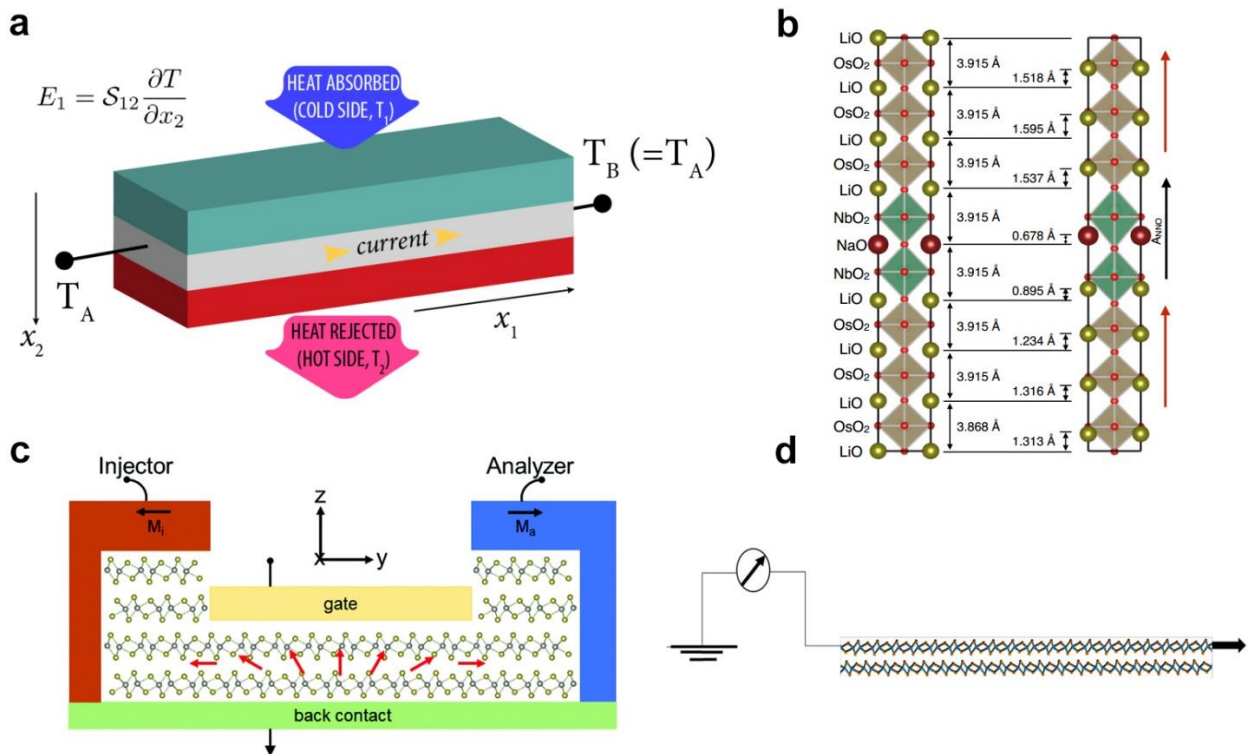

**Fig. 9.** Potential applications of ferroelectric/polar metals. (a) $(Sr,Ca)Ru_2O_6$-based anisotropic thermoelectric device. The temperature gradient along the $x_2$ direction results in an electric field along the $x_1$ direction, which can generate a current density. Note that along the $x_1$ direction, there is no temperature gradient, i.e., $\frac{\partial T}{\partial x_1} = 0$, and thus $T_A = T_B$. Adapted by permission from Springer Nature Customer Service Centre GmbH: Springer Nature, Nature Communications, ref. 30, Copyright (2014)[30]. (b) Crystal structure of the centrosymmetric nanocapacitor (left) and the relaxed equilibrium structure of the ferroelectric nanocapacitor (right) [LiO-$(OsO_2$-LiO$)_n$/$NbO_2$-$(NaO$-$NbO_2)_m$] ($m$=1, $n$=6) with insulating $NaNbO_3$ ($m$=1) between $LiOsO_3$ electrodes ($n$=6). The direction of the polar displacements in the electrodes and the ferroelectric film ($A_{NNO}$) are indicated with arrows. Reprinted from ref. 122, with the permission of AIP Publishing[122]. (c) Struture of a $WTe_2$-based spin field effect transistor (spin-FET). Injector and analyzer are ferromagnetic electrodes with in-plane and out-of-plane magnetic anisotropies, respectively, while the gate and back electrodes allow the modulation/switch of ferroelectric polarization of bilayer $WTe_2$. The red arrows represent the spin direction when electrons travel from the injector to the analyzer. Republished with permission of Royal Society of Chemistry, from ref. 101; permission conveyed through Copyright Clearance Center, Inc[101]. (d) A model of nanogenerator based on $WTe_2$ bilayer. Reprinted (adapted) with permission from ref. 100. Copyright (2018) American Chemical Society[100].